\documentclass[11pt,a4paper]{article}
\usepackage{amsmath}
\usepackage{amsthm}
\usepackage{amssymb}
\usepackage{graphicx}

\begin{document}

\newcommand{\beq}[1]{\begin{equation}\label{#1}}
\newcommand{\eq}{\end{equation}}

\newtheorem{predl}{Proposition}
\newtheorem{lem}{Lemma}[section]
\newtheorem{theor}{Theorem}[section]

\begin{center}

\vspace{20mm}

{\Large{\sc Explicit Semi-invariants and Integrals of the Full Symmetric $\mathfrak{sl}_n$ Toda Lattice}
}\\


\vspace{35pt}
{\sc Yu.B. Chernyakov${}^{\; a}$ and A.S. Sorin$^{\; b}$}\\[15pt]

{${}^a$\small Institute for Theoretical and Experimental Physics\\ Bolshaya Cheremushkinskaya, 25, 117218 Moscow, RUSSIA and\\ Bogoliubov Laboratory of Theoretical Physics\\ Joint Institute for Nuclear Research\\ 141980 Dubna, Moscow region,  RUSSIA.\\}e-mail: {\small \it
chernyakov@itep.ru}\vspace{10pt}

{${}^b$\small Bogoliubov Laboratory of Theoretical Physics and\\
Veksler and Baldin Laboratory of High Energy Physics\\
Joint Institute for Nuclear Research \\ 141980 Dubna, Moscow Region, RUSSIA
\\ }e-mail: {\small \it
sorin@theor.jinr.ru}\vspace{10pt}

\vspace{3mm}

\end{center}

\begin{abstract}
We show how to construct semi-invariants and integrals of the full
symmetric $\mathfrak{sl}_n$ Toda lattice for all n. Using the Toda equations for the
Lax eigenvector matrix we prove the existence of semi-invariants which
are homogeneous coordinates in the corresponding projective spaces. Then
we use these semi-invariants to construct the integrals.
The existence of additional integrals which constitute a full set of
independent non-involutive integrals was known but the chopping and
Kostant procedures have crucial computational complexities already for
low-rank Lax matrices and are practically not applicable for higher
ranks. Our new approach solves this problem and results in simple
explicit formulae for the full set of independent semi-invariants and
integrals expressed in terms of the Lax matrix and its eigenvectors, and
of eigenvalue matrices for the full symmetric $\mathfrak{sl}_n$ Toda lattice.
\end{abstract}


\section{Introduction} The iso-spectral flows of the full symmetric $\mathfrak{sl}_n$ Toda lattice
\beq{LM}
\frac{\partial}{\partial t_{d-1}}{L} = [B_{d},L], \ \ d=\overline{2,n}
\eq
with the symmetric Lax operator $L$ ($L^T=L$) and antisymmetric evolution operator $B_d$ ($B_d^T=-B_d$)
\beq{Lax} L \equiv \left(
\begin{array}{c c c c c c}
 a_{11} & a_{12} & ... & a_{1n}\\
 a_{12} & a_{22} & ... & a_{2n}\\
 ... & ... & ... & ...\\
 a_{1n} & a_{2n} & ... & a_{nn}\\
\end{array}
\right), \ \
B_{d} \equiv (L^{d-1})_{>0} - (L^{d-1})_{<0}
\eq
are the compatibility conditions for the linear system
\beq{compat-Lax}
\left\{
\begin{array}{c}
L \Psi = \Psi \Lambda,\\
\ \\
\frac{\partial}{\partial t_{d-1}} \Psi = B_{d} \Psi,
\end{array}
\right.
\eq
where $\Psi\in SO(n)$ is the eigenvector matrix and $\Lambda\equiv diag(\lambda_1,...,\lambda_n)$ is the corresponding eigenvalue matrix
(see, e.g. \cite{A, Ad, K1, S, DLNT} and references therein).
The full symmetric $\mathfrak{sl}_n$ Toda lattice is a Hamiltonian system: its phase space is the coadjoint-action orbit of the Borel subgroup $B^+_n \in SL_n(\mathbb R)$ and the corresponding iso-spectral Hamiltonians are
\beq{IsoH}
 H_d=\frac{1}{d} Tr L^{d}, \ d=\overline{2,n}.
\eq
Such a phase space can be realized in two different ways by identifying the algebra $\mathfrak{sl}_n$ with its dual using the $\mathfrak{sl}_n$ Killing form. The first method is based on the decomposition $\mathfrak{sl}_n=\mathfrak{n}^{-}_{n} \oplus \mathfrak{b}^+_n$, so that the phase space can be identified with the coadjoint-action orbit of the Borel subgroup $B^+_n$ in the affine space $\mathfrak{b}^-_n + \epsilon$, where $\epsilon$ is the sum of $\mathfrak{sl}_n$ simple roots, $\mathfrak{b}^+_n$ and $\mathfrak{b}^-_n$ are the algebras of upper and lower triangular matrices, respectively, and $\mathfrak{n}^{-}_n$ is the algebra of strictly lower triangular matrices. In the present letter we consider the other realization based on the decomposition $\mathfrak{sl}_n=\mathfrak{so}_n \oplus \mathfrak{b}^+_n$, so that one can identify the space of symmetric matrices $Symm_n$ with the dual space of the Lie algebra of the Borel subgroup, $Symm_n\cong(\mathfrak{b}_n^+)^*$, hence introduce a symplectic structure on $Symm_n$ pulling it back from $(\mathfrak{b}_n^+)^*$. Based on these two approaches one gets two different integrable systems called the full $\mathfrak{sl}_n$ Kostant-Toda lattice and the full symmetric $\mathfrak{sl}_n$ Toda lattice.

The dimension of the phase space of these systems is $2[\frac{1}{4} n^{2}]$ which is greater than $2n$, and in order to show their Liouville integrability an involutive set of $[\frac{1}{4} n^{2}]$ integrals is required, so that it is not enough to have only $n$ iso-spectral integrals (\ref{IsoH}). Such a set can be derived by the chopping procedure \cite{DLNT}. The two different families of the integrals in involution of the full $\mathfrak{sl}_n$ Kostant-Toda lattice were observed at $n=4$ in \cite{EFS,S2}. A formula for the integral derivation without using the chopping procedure is discussed in the recent paper \cite{T}, in which it is used for a generalization to the quantum case.

Besides the above-mentioned involutive integrals, the Toda systems possess non-involutive integrals as well, so that the Toda systems are integrable in the non-commutative sense \cite{Neh}. The extension of the set of involutive integrals by non-involutive ones as well as construction of semi-invariants was explored in \cite{A, BG-1, GS, BG-2, FS, Fre:2011uy}. A construction procedure of the full set of non-involutive independent integrals of the Toda systems for classical Lie algebras is described in \cite{BG-1, GS, BG-2}, where it is based on the Kostant decomposition of a generic Lie algebra element of a maximally split simple Lie algebra (see \cite{BG-1, GS, BG-2} for details). This procedure is so complicated technically, that one can actually apply it only to few algebras of smallest ranks.

In the present letter which is the short version of our paper \cite{CS} we show how to obtain the explicit formulae for the full set of independent semi-invariants and integrals of the full symmetric $\mathfrak{sl}_n$ Toda lattice, expressed both in terms of the Lax operator $L$ and its $SO(n)$ eigenvector matrix $\Psi$.

\section{Properties of the full symmetric $\mathfrak{sl}_n$ Toda lattice}
\subsection{Chopping procedure in the full symmetric $\mathfrak{sl}_n$ Toda lattice \cite{DLNT}}
The $[\frac{n^{2}}{4}]$ functionally independent integrals in involution for the full symmetric $\mathfrak{sl}_n$ Toda lattice on the generic orbit of $2[\frac{n^{2}}{4}]$-dimension, which provide its Liouville integrability,
\beq{Integrals}
I_{m,k} = \frac{E_{m,k}}{E_{0,k}}, \ \ \ 0 \leq k \leq \Big[\frac{n-1}{2}\Big],  \ \ \ 1 \leq m \leq n-2k
\eq
are expressed in terms of the coefficient-functions $E_{m,k}$ of the characteristic polynomials:
\beq{Pkn}
\begin{array}{c}
\det(L - \mu I)_{k} = \sum_{m=0}^{n-2k} E_{m,k}\, \mu^{n-2k-m}, \,\,\, 0 \leq k \leq [n/2]\ ,
\end{array}
\eq
where $(L - \mu I)_{k}$ is the $(n-k)\times (n-k)$-matrix derived by chopping $k$ upper rows and $k$ right column of the matrix $(L - \mu I)$; $[ \ \ \ ]$ means the integer part. The integrals $I_{m,k}$ are Casimirs at $m=1$, their number is $[\frac{n-1}{2}]$.

\subsection{Dynamics of minors of the eigenvector matrix $\Psi$}
Now, using eqs. (\ref{compat-Lax}) we derive explicitly the iso-spectral flows of the special minors of $\Psi$, which is used in what follows to derive the semi-invariants and integrals.

\begin{predl}
The iso-spectral flows of the minors $M_{\frac{1,2,...,k}{i_{1},i_{2},...,i_{k}}}$  $\Big({M}_{\frac{n-k+1,...,n}{i_{1},i_{2},...,i_{k}}}\Big)$ of $\Psi$, obtained by the intersection of its first (last) $k$ rows $1,2,...,k$ ($n-k+1,...,n$) with any set of $k$ columns $i_{1},i_{2},...,i_{k}$, are
\beq{minors-d}
\begin{array}{c}
\frac{\partial}{\partial t_{d-1}}M_{\frac{1,2,...,k}{i_{1},i_{2},...,i_{k}}} = \Big(-\sum_{j=1}^{k}a^{(d-1)}_{jj} + \sum_{i_{m}=i_{1}}^{i_{k}}\lambda^{d-1}_{i_{m}}\Big)\, M_{\frac{1,2,...,k}{i_{1},i_{2},...,i_{k}}} \ ,\\
\frac{\partial}{\partial t_{d-1}}M_{\frac{n-k+1,...,n}{i_{1},i_{2},...,i_{k}}} ~= \Big(+\sum_{j=n-k}^{n}a^{(d-1)}_{jj} - \sum_{i_{m}=i_{1}}^{i_{k}}\lambda^{d-1}_{i_{m}}\Big)\, M_{\frac{n-k+1,...,n}{i_{1},i_{2},...,i_{k}}}
\end{array}
\eq
where $a^{(d)}_{ij}$ are the matrix elements of the operator $L^{d}$, in particular, these minors are semi-invariants of the iso-spectral flows.
\end{predl}

The proof of this proposition is based on straightforward calculations of dynamics of the minors of $M$ and on induction.

\section{Structure of independent non-involutive integrals}
The number of independent non-involutive integrals without taking into account the Casimirs \cite{BG-1, GS}
\beq{Integrals-number}
\begin{array}{c}
N_{n} = \frac{n(n-1)}{2} - [\frac{n+1}{2}] + 1
\end{array}
\eq
consists of three different contributions
\beq{IntContr}
\begin{array}{c}
N_{n} \equiv N_{n}^{Iso} + N_{n}^{Chopp} + N_{n}^{Add}
\end{array}
\eq
which come from the iso-spectral integrals $N_{n}^{Iso}$, the integrals $N_{n}^{Chopp}$ derived by the chopping procedure with exclusion of the Casimirs, and the additional integrals $N_{n}^{Add}$ which extend the above--mentioned two families to a full set of the integrals, respectively. The sum of the iso-spectral integrals and the integrals obtained by the chopping procedure, which ensures the Liouville integrability, is \cite{DLNT}
\beq{IntChop}
\begin{array}{c}
N_{n}^{Iso} + N_{n}^{Chopp}=[\frac{n^{2}}{4}] \ .
\end{array}
\eq
From equalities (\ref{Integrals-number}), (\ref{IntContr}) and (\ref{IntChop}) one can obviously conclude that
\beq{IntChopAddEq}
\begin{array}{c}
N_{n}^{Add}=N_{n}^{Chopp} \equiv [\frac{(n-2)^{2}}{4}].
\end{array}
\eq

The structure of the special minors of $L^k \ (k \in \mathbb{N})$ shows that they are quadratic expressions of the special
minors of $\Psi$, that appear in eqs. (\ref{minors-d}) which are semi-invariants. In this way we establish that the former minors are
semi-invariants as well, and construct invariants using these semi-invariants:
\begin{predl}
The minors $A^{(k)}_{\frac{n-m+1,...,n}{1,2,...,m}}$ of the matrices $L^{k}$ can be represented as
\beq{A-lambda-psi}
A^{(k)}_{\frac{n-m+1,...,n}{1,2,...,m}}=\sum_{i_{1},i_{2},...,i_{m}} \lambda^{k}_{i_{1}}\lambda^{k}_{i_{2}} \cdot...\cdot \lambda^{k}_{i_{m}} M_{\frac{1,2,...,m}{i_{1},i_{2},...,i_{m}}}M_{\frac{n-m+1,...,n}{i_{1},i_{2},...,i_{m}}}\ ,\\
\eq
so that they are the semi-invariants which generate the invariants of the iso-spectral flows:
\beq{A-lambda-psi-Integrals}
J_{k_{1},k_{2}}= \frac{A^{(k_{1})}_{\frac{n-m+1,...,n}{1,2,...,m}}}{A^{(k_{2})}_{\frac{n-m+1,...,n}{1,2,...,m}}} \ \ \ \ \ \ \ (k_{1}, \, k_{2} \in \mathbb{N}) \ .
\eq
\end{predl}
The proof of eq. (\ref{A-lambda-psi}) is based on the decomposition of the minors $A_{\frac{n-m+1,...,n}{1,2,...,m}}$ in terms of matrix elements of the Lax operator $L$, then its decomposition in terms of $\lambda, \ \psi$ and regrouping the result. The proof of eq. (\ref{A-lambda-psi-Integrals}) is based on eqs. (\ref{minors-d}) and (\ref{A-lambda-psi}).

An interesting fact is that the integrals obtained by the chopping procedure have the same structure as the integrals (\ref{A-lambda-psi-Integrals}) which are rational functions of a bilinear product of the $\Psi$-special minors:
\begin{predl}
The involutive integrals (\ref{Integrals}) derived by the chopping procedure can be represented as
$$
I_{m,k} = \frac{\sum_{i_{1}<...<i_{s}<...<i_{k}} \lambda_{i_{1}} \cdot ... \cdot \lambda_{i_{k}} \sum_{j_{1}<...<j_{t}<...<j_{m}, \ j_{t} \neq i_{s}} \lambda_{j_{1}} \cdot ... \cdot \lambda_{j_{m}} M_{\frac{1,...,k}{i_{1},...,i_{k}}}M_{\frac{n-k+1,...,n}{i_{1},...,i_{k}}}}{\sum_{l_{1}<...<l_{r}<...l_{k}} \lambda_{l_{1}} \cdot...\cdot \lambda_{l_{k}} M_{\frac{1,...,k}{l_{1},...,l_{k}}}M_{\frac{n-k+1,...,n}{l_{1},...,l_{k}}}},$$
\beq{Integrals-Plukk}
\begin{array}{c}
1 \leq s,r \leq k, \ 1 \leq t \leq m, \ 1 \leq i_{s},j_{t},l_{r} \leq n\\
\ \\
0 \leq k \leq [\frac{n-1}{2}],  \ \ \ 1 \leq m \leq n-2k.
\end{array}
\eq
\end{predl}
The proof of this proposition is similar to the proof of Proposition 2.
Note that it is possible to express the integrals (\ref{Integrals-Plukk}) via the integrals (\ref{A-lambda-psi-Integrals}) using eqs. (\ref{A-lambda-psi}), which allow to express all products $M_{\frac{1,2,...,k}{i_{1},i_{2},...,i_{k}}}M_{\frac{n-k+1,...,n}{i_{1},i_{2},...,i_{k}}}$ entering into eq. (\ref{Integrals-Plukk}) in terms of  $A^{(k)}_{\frac{n-m+1,...,n}{1,2,...,m}}$.

The functions from the set (\ref{A-lambda-psi-Integrals}) are not all independent, but this set comprises all independent non-involutive integrals of the full symmetric $\mathfrak{sl}_n$ Toda lattice. In order to evaluate their number and describe them, we first consider the Pl\"ucker embedding which describes the flag space $FL_{n}(\mathbb{R})=SO(n, \mathbb{R})/S$, where the discrete subgroup $S$ of the group $SO(n, \mathbb{R})$ consists of the diagonal matrices with the entries $\pm 1$, in terms of the coordinates
\beq{Plukk-1}
X_{i_{1},i_{2},...,i_{m}} = M_{\frac{1,2,...,m}{i_{1},i_{2},...,i_{m}}}(\Psi)\ , \ \ \Psi\in SO(n,\,\mathbb R)
\eq
modulo the quadratic relations \cite{F}
\beq{Quadratic-Relation}
\begin{array}{c}
X_{i_{1},i_{2},...,i_{m_{1}}} \cdot X_{j_{1},j_{2},...,j_{m_{2}}}-\sum X_{i^{'}_{1},i^{'}_{2},...,i^{'}_{m_{1}}} \cdot X_{j^{'}_{1},j^{'}_{2},...,j^{'}_{m_{2}}}=0 \ , \ \ \  m_{1} \leq m_{2}
\end{array}
\eq
where the sum is taken over all pairs of multi-indices obtained by interchanging the first $k$ subscripts of $j_{1},j_{2},...,j_{m_{2}}$ with $k$ subscripts of $i_{1},i_{2},...,i_{m_{1}}$, maintaining the order. It follows from (\ref{A-lambda-psi-Integrals}--\ref{Integrals-Plukk}) that the integrals are rational functions of $\lambda$ and $\psi$. The numerators and the denominators of these functions consist of the sums such that each of their terms is the product of a polynomial function of $\lambda$ and the function $\varphi(M(\psi))$ defined by:
\beq{CrossRatio}
\varphi(M(\Psi))=\frac{M_{\frac{1,2,...,m}{i_{1},i_{2},...,i_{m}}}M_{\frac{n-m+1,...,n}{i_{1},i_{2},...,i_{m}}}}{M_{\frac{1,2,...,m}{j_{1},j_{2},...,j_{m}}}M_{\frac{n-m+1,...,n}{j_{1},j_{2},...,j_{m}}}},
\eq
Taking into account (\ref{minors-d}) one gets
\begin{predl}
The functions $\ \varphi(M(\psi))$ (\ref{CrossRatio}) are integrals with respect to the one-parametric iso-spectral flows. The number of independent integrals in the set (\ref{CrossRatio}) is
\beq{FLdim}
N_{\Psi} = dim Fl_{n}(\mathbb{R}) - (n-1)\equiv \frac{n(n-1)}{2} - (n-1) \ .
\eq
\end{predl}

The proof of this proposition is based on the direct calculation of the
$\varphi(M(\psi))$--dynamics using eqs. (\ref{minors-d}) and taking into account that $\varphi(M(\psi))$ are the functions on the flag space defined modulo the iso-spectral flows in a generic point.

In order to evaluate the total number of independent integrals in the set (\ref{A-lambda-psi-Integrals}) it is necessary to subtract the number of Casimirs $N_{Cas}=[\frac{n-1}{2}]$ from the $N_{\Psi}$ (\ref{FLdim}) and add the number of the iso-spectral integrals $N_{Iso}=n-1$ to the result, i.e.
\beq{formulas-fl-2}
\begin{array}{c}
N_{\Psi}-N_{Cas}+N_{Iso}\equiv  \frac{n(n-1)}{2} - [\frac{n+1}{2}] + 1 \ ,
\end{array}
\eq
which precisely reproduces the total number $N_n$ (\ref{Integrals-number}) of independent non-involutive integrals of the full symmetric $\mathfrak{sl}_n$ Toda lattice.

\section{Conclusions}
Now we are ready to describe explicitly the full non-involutive set of independent integrals expressed in terms of matrix elements of the Lax operator $L$. For this goal one can choose the following integrals from the set (\ref{A-lambda-psi-Integrals}):
\beq{A-lambda-psi-Integrals-2}
\begin{array}{c}
J_{k,1}= \frac{A^{(k)}_{\frac{n-m+1,...,n}{1,2,...,m}}}{A_{\frac{n-m+1,...,n}{1,2,...,m}}} \ \ \ \mbox{with}\\
m=1\ , \  \ k=\overline{3,n-1} \  \ \ \ \ \ \ \ \ \ \ \ \ (\# \ \mbox{integrals} = n-3) \\
 2 \leq m < [\frac{n}{2}]\ , \ k=\overline{2,n-2m} \ \ \ (\# \ \mbox{integrals} = [\frac{(n-4)^{2}}{4}])\\
 m=[\frac{n}{2}]\ , \ \ k=\overline{2,[\frac{(n-2)^2}{4}]+1} \ \ \ \ \ (\# \ \mbox{integrals} = [\frac{(n-2)^2}{4}])
\end{array}
\eq
together with the $(n-1)$ iso-spectral integrals $H_d$ (\ref{IsoH}), so that the total number of the integrals is given by eq. (\ref{Integrals-number}).

In order to form the full non-involutive set of independent integrals expressed in terms of the eigenvector matrix $\Psi$, one should for every $m$ from the range $1 \leq m < [\frac{n}{2}]$ select $n-2m-1$ integrals from the set (\ref{CrossRatio}) and for $m=[\frac{n}{2}]$ select $[\frac{(n-2)^2}{4}]$ integrals from the set (\ref{CrossRatio}). The union of these integrals together with the $n-1$ integrals expressed in terms of the eigenvalue matrix $\Lambda$ gives a full non-involutive set of independent integrals of the full symmetric $\mathfrak{sl}_n$ Toda lattice, which number is $N_{n}$ (\ref{Integrals-number}).

In this paper we developed a new approach to derive integrals of motion of the full symmetric $\mathfrak{sl}_n$ Toda lattice which uncovers its "genetics" from the viewpoint of flag spaces. We use the semi-invariants (\ref{minors-d}), which are Pl\"ucker coordinates (\ref{Plukk-1}) in the corresponding projective spaces, in order to construct explicitly the full set of the non-involutive integrals expressed both in terms of the Lax matrix (\ref{A-lambda-psi-Integrals-2}) and its eigenvalue and eigenvector matrices (\ref{CrossRatio}) of arbitrary ranks. \\
Our approach is much simpler than the one based on Kostant procedure \cite{BG-1,GS,BG-2} and avoids the crucial computational complexities
appearing in the latter procedure even for low-rank Lax matrices, which prevent it use for the higher ranks. \\
The simplicity of the advocated approach is exemplified by the additional integral $J={A^{(2)}_{\frac{3,4}{1,2}}}(A_{\frac{3,4}{1,2}})^{-1}$ of the full $\mathfrak{sl}_4$ Kostant-Toda lattice. In order to derive it the authors of \cite{EFS,S2} applied the isomorphism $\mathfrak{sl}_4
\leftrightarrow \mathfrak{so}_6$ and the $\mathfrak{so}_6$-chopping procedure.

The results of the present paper are crucial to establish the Bruhat order in the full symmetric $\mathfrak{sl}_n$ Toda lattice \cite{CSS}.

Detailed proofs of the formulae of the Propositions are given in \cite{CS}. The technique that we have developed in the present paper has further extensions and applications. The generalization to the Toda lattices defined for other Lie algebras and homogeneous spaces will be given elsewhere.

\paragraph{Acknowledgments}
The authors would like to thank G.I. Sharygin and D. Sternheimer for fruitful discussions and remarks.
The work of Yu.B. Chernyakov was supported in part by grants RFBR-12-02-00594 and by the Federal Agency for Science and Innovations of Russian Federation under contract 14.740.11.0347. The work of A.S. Sorin was supported in part by the RFBR Grants No. 11-02-01335-a, No. 13-02-91330-NNIO-à and No. 13-02-90602-Arm-a.


\begin{thebibliography}{60}
\bibitem{A}
A.A. Arhangelskii, Completely integrable hamiltonian systems on a group of triangular matrices, Mathematics of the USSR-Sbornik, 36:1, 127 -- 134 (1980).

\bibitem{Ad}
M. Adler, On a trace functional for pseudo-differential operators and the symplectic structure of the Korteweg-de Vries equation, Invent. Math., 50, 219 -- 248 (1979).

\bibitem{K1}
B. Kostant, The solution to a generalized Toda lattice and representation theory, Adv. in Math. 34, 195 -- 338 (1979).

\bibitem{S}
W. W. Symes, Systems of Toda type, inverse spectral problems, and representation theory, Invent. Math. 59, no. 1, 13 -- 51 (1980).

\bibitem{DLNT}
P. Deift, L. C. Li, T. Nanda, and C. Tomei, The Toda flow on a generic orbit is integrable, CPAM 39, 183 -- 232 (1986).

\bibitem{EFS}
N. Ercolani, H. Flaschka, and S. Singer, The geometry of the full Kostant-Toda lattice In: Integrable Systems,
Vol. 115 of Progress in Mathematics, Birkhauser, 181 -- 226 (1993).

\bibitem{S2}
B. A. Shipman, The geometry of the full Kostant-Toda lattice of sl(4;C), Journal of Geometry and Physics
33, 295 -- 325 (2000).

\bibitem{T}
D.Talalaev, Quantum generic Toda system, arXiv:1012.3296.

\bibitem{Neh}
N.N. Nehoroshev, Action-angle variables and their generalization,  Tr. Mosk. Mat. O.-va. 26, 181 -- 198 (1972).

\bibitem{BG-1}
A. M. Bloch and M. Gekhtman, Hamiltonian and gradient structures in the Toda flows, J. Geom. Phys. 27, 230 -- 248 (1998).

\bibitem{GS}
M. Gekhtman and M. Shapiro, Noncommutative and commutative integrability of generic Toda flows in simple Lie algebras, Comm. Pure and Appl. Math. 52, 53 -- 84 (1999).

\bibitem{BG-2}
A. M. Bloch and M. Gekhtman, Lie algebraic aspects of the finite nonperiodic Toda
flows, J. Comp. Appl. Math. 202, 3 -- 25 (2007).

\bibitem{FS}
P. Fre, A.S. Sorin, The arrow of time and the Weyl group: all supergravity billiards are integrable, Nucl. Phys., B 815 (2009), 430 [arXiv:0710.1059 [hep-th]].

\bibitem{Fre:2011uy}
  P.~Fre, A.~S.~Sorin and M.~Trigiante,
  Integrability of Supergravity Black Holes and New Tensor Classifiers of Regular and Nilpotent Orbits,
  JHEP {\bf 1204} (2012) 015
  [arXiv:1103.0848 [hep-th]].

\bibitem{F}
W. Fulton, Young Tableaux, Cambridge University Press, 1977.

\bibitem{CSS}
  Y.~B.~Chernyakov, G.~I.~Sharygin and A.~S.~Sorin,
  Bruhat Order in Full Symmetric Toda System,
  arXiv:1212.4803 [nlin.SI].

\bibitem{CS}
Y.~B.~Chernyakov, A.~S.~Sorin,
 Semi-invariants and Integrals of the Full Symmetric $\mathfrak{sl}_n$ Toda Lattice,
  arXiv:1312.4555 [nlin.SI].


\end{thebibliography}
\end{document}